# Serial Parallel Reliability Redundancy Allocation Optimization for Energy Efficient and Fault Tolerant Cloud Computing

Gutha Jaya Krishna
Administrative Staff College of India
Hyderabad, India
krishna.gutha@gmail.com

*Abstract*— Serial-parallel redundancy is a reliable way to ensure service and systems will be available in cloud computing. That method involves making copies of the same system or program, with only one remaining active. When an error occurs, the inactive copy can step in as a backup right away, this provides continuous performance and uninterrupted operation. This approach is called parallel redundancy, otherwise known as active-active redundancy, and it's exceptional when it comes to strategy. It creates duplicates of a system or service that are all running at once. By doing this fault tolerance increases since if one copy fails, the workload can be distributed across any replica that's functioning properly. Reliability allocation depends on features in a system and the availability and fault tolerance you want from it. Serial redundancy or parallel redundancies can be applied to increase the dependability of systems and services. To demonstrate how well this concept works, we looked into fixed serial parallel reliability redundancy allocation issues followed by using an innovative hybrid optimization technique to find the best possible allocation for peak dependability. We then measured our findings against other research.

## I. INTRODUCTION

Cloud computing relies on reliability optimization to guarantee that users have uninterrupted, high-performing access to systems and services. When services fail or things go wrong, the methods that are used to keep systems running and minimize damage if they can't are load balancing, redundancy, and fault tolerance. Redundancy is especially important in maintaining resilience as it keeps everything going in times of failure. Depending on the thing's unique needs, there are a few areas that this can be achieved: hardware, software, or data redundancies. But no matter how it's done, having a plan is always important. With the inevitability of failures and malfunctions, you should use redundancy as it will reduce interruptions and increase availability, making sure both performance and functionality.

Serial and parallel redundancy are great aids to cloud systems. Especially when it comes to availability, having a failover system with serial redundancy is key. In case the primary cloud fails, users can automatically be moved over to the backup one – guaranteeing that the service remains live for all users. By setting up this kind of redundancy we can really enhance security and performance in your cloud system. Faster recovery from incidents and easier maintenance during work can be achieved by allowing one system to stay active while we perform work on the other. Parallel redundancy not only ensures that your cloud system stays online even if a single one fails but also helps take on any amount of workload with ease. This does so by sharing the load among multiple systems and enables horizontal scaling and guarantees unparalleled performance levels no matter how much traffic increases. Using parallel redundancy will strengthen our security by providing us with the ability to segment each component from one another. This will help us mitigate potential risks that could affect the integrity of an entire system in an instant.

To make sure systems can still work even if they fail or break down, organizations can take advantage of serial and parallel redundancy. This gives companies the ability to reduce the impact of unexpected problems and to maintain reliability. Business owners have the ability to use this by simply splitting up tasks with duplicates, and if need be, with auto-scaling too. Through all of this, failover capabilities and load balancing also play a role in keeping performance levels steady when something does fail. In conclusion, without one of our safeguards our systems would easily fold under pressure. But by using both serial and parallel redundancy we can improve multiple layers of protection to not only keep everything running smooth but also ensure no individual components hold us back from improving on that experience through all likelihoods.

Serial Parallel Reliability Redundancy Allocation (SPRRA) Optimization helps in the following tasks for achieving energy efficiency:

- Optimal Resource Utilization: The most energy-efficient setup of resources that can maintain the desired reliability level is shown through SPRR optimization. This means that it identifies the best way to place energy-efficient components in parallel so that reliability isn't compromised in exchange for energy savings.

- Load-Adaptive Operation: Saving energy when there's a low demand is what the optimization model does by dynamically adjusting resources (CPUs, storage, and network bandwidth). For example, virtual machines or containers can be turned on or off if needed for the work load at hand, no longer will you ever have to worry about wasting electricity on idle resources.

- Intelligent Power Management: There's a time where redundancy isn't needed right away and intelligent algorithms can pick up on it and switch components to low-power states. During non-peak hours, non-critical parallel components can be powered down while maintaining system reliability with serial components.

Serial Parallel Reliability Redundancy Allocation (SPRRA) Optimization helps in the following tasks for achieving fault tolerance:

- Redundancy Allocation: You improve your system's resilience from faults by strategically placing redundancy where it's most impactful with SPRR optimization. By doing so you'll be able to manage higher failure rates in parallel configurations before they cause an impact.

- Enhanced Availability: Regardless of component failures, a calculated serial-parallel arrangement ensures that there are always active and available resources to handle user requests.

- Graceful Degradation: When experiencing failure, instead of going through a complete shutdown the system continues operating at a reduced capacity therefore allowing repairs to be made while still being functional.

The paper is organized as follows: Motivation is given in Section II, followed by overall advantages of the work in the Section II. Related works are presented in Section IV. An example and a sample problem definition are given in Section V. Proposed methodology and results are presented in Sections VI and VII. Finally in Sections IX and X, research questions, future directions and conclusions are presented.

## II. MOTIVATIONS

In cloud computing, Serial Parallel Reliability-Redundancy Allocation (SPRRA) is used to improve the reliability and availability of systems and services. Systems in the cloud are usually complex and distributed with a chance of many failures that can lead to service outages and downtime. SPRRA allocates redundancy in a way that reduces failure risk and overall improves cloud system performance.

SPRRA takes the best of both worlds, by using both serial and parallel redundancy. Serial redundancy is pretty easy to understand, you put a lot of devices in series (basically on top of each other). If one stops working then the whole system shuts down. Parallel is when all of the devices work together and not really off each other. Combining these two techniques can minimize the risk of service failures and maximize the availability for cloud systems.

For starters, cloud computing is expensive. Especially when it comes to redundancy allocation. But then why do cloud providers continue to do it? It's simple, with the big price tag comes benefits like better availability and reliability. And that's what SPRRA does. This method of allocating redundancy has one goal in mind to minimize cost while still providing the desired level of reliability and availability.

Ultimately, improving reliability and availability of cloud systems and services while keeping costs as low as possible is the motivation behind SPRRA in cloud computing.

## III. OVERALL ADVANTAGES OF THE WORK

- Increased Reliability and Availability: SPRRA can improve the availability and reliability of services by removing the risk of service failures and downtime.

- Optimal Redundancy Allocation: The cost of redundancy and increased reliability can balance each other out by using SPRRA to figure out the best way to allocate it.

- Improved fault tolerance: Combining serial and parallel redundancy makes SPRRA an effective option for maintaining system performance. No matter what, it's efficient.

- Reduced maintenance costs: This new system allows cloud systems to conduct maintenance and repairs with minimal disruption. On top of that, it also reduces the overall cost.

- Scalability: It doesn't matter what type of cloud system you have. SPRRA is flexible enough to be scaled for anyone's needs.

- Enhanced customer satisfaction: Through enhancing these things, businesses can see increased satisfaction among customers that leads to retention and more revenue.

## IV. RELATED WORKS

There is a significant amount of literature [1], [2] on serial and parallel reliability redundancy allocation [3]–[40]. Improved Non-equilibrium Simulated Annealing (INESA) [41] was developed for this problem. Different optimization techniques were applied to solve the problem. In particular, threshold accepting [42], a variant of simulated annealing (SA), Ant Colony Optimization (ACO) [43], the Modified Great Deluge Algorithm (MGDA) [44], Improved Modified Harmony Search (IMHS) [45] coupled with Modified Differential Evolution (MDE) [46] were applied to solve the problem. Each subsequent technique was found to produce superior results compared to the previous one.

## V. EXAMPLE AND PROBLEM

Here's an example problem for Serial Parallel Reliability-Redundancy Allocation (SPRRA) in cloud computing:

A cloud service provider offering a service with high availability. The agreement states that you need to maintain 99.999% uptime per year. However, the system is made up of 5 components each with a reliability of 0.9995 and you want to figure out how many redundant parts you need to keep everything running while keeping the cost down. With SPRRA, business or industry can calculate the redundant components needed for the Service Level Agreement (SLA).

Let's say we use SPRRA to calculate the amount of redundancy we'd need for it to hit the SLA. Now if you know how many are needed for series and parallel, finding out what the total cost of redundancy is is simple. To do this problem we can utilize SPRRA to get a good allocation for redundancy in the system. Here are some steps:

1. Calculate reliability without redundancy: $(0.9995)^5 = 0.9975$.

2. Calculate reliability with redundancy: 0.99999.

3. Figure out how many components to use: We can do this by using series and parallel redundancy, figures n as series and m as parallel then apply this formula: $R = (0.9995^n) * (1 - (1 - 0.9995)^m)$. Once we get a reliable number then it'll be time to move on.

4. Calculate your total costs: This one's simple but it's worth doing it again manually just as reassurance. The total cost is calculated by multiplying your costs per component by how many you have in series/parallel: $1000 * m + n. Doing this will give us a clear picture of how much we need to allocate. So far we've been talking in hypotheticals but now it's time to put it into action. In this example let's use n = 2, and m = 3. Now we know the optimal allocation is to use 2 redundant components in series and 3 in parallel, at a total cost of $5000.

Overall, optimization if employed instead of trial and error can provide many benefits for cloud computing, improving resource utilization, system performance, reliability, and availability, while reducing costs and enhancing the user experience. By applying optimization techniques, cloud providers can ensure that their systems are efficient, effective, and responsive to the needs of their users.

A standard serial-parallel system illustration is given in Fig. 1, which is considered for the reliability redundancy allocation problem (RRAP).

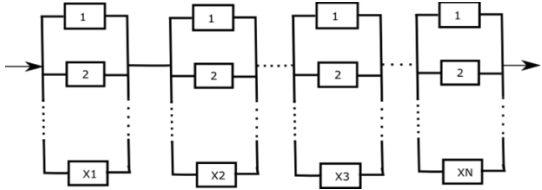

Fig. 1. Serial-Parallel System.

The standard serial-parallel RRAP considered for reliability optimization is a mathematical integer optimization problem given below [45], where the objective is to maximize reliability $R_s$ subject to satisfying the constraints $g_1$ and $g_2$. The constants of reliability, cost, and weight required for solving the problem are presented in Table I.

Find the optimal $x_i$, i=1, 2, ..., 15

$$\text{Max } R_s = \prod_{i=1}^{N=15} [1 - (1 - R_i)^{x_i}]$$

Subject to:

$$g_1 = \sum_{i=1}^{N=15} C_i x_i \leq 400$$

$$g_2 = \sum_{i=1}^{N=15} W_i x_i \leq 414$$

TABLE I. CONSTANTS

| i | 1 | 2 | 3 | 4 | 5 | 6 | 7 | 8 | 9 | 10 | 11 | 12 | 13 | 14 | 15 |
|---|---|---|---|---|---|---|---|---|---|----|----|----|----|----|----|
| $R_i$ | 0.9 | 0.75 | 0.65 | 0.80 | 0.85 | 0.93 | 0.78 | 0.66 | 0.78 | 0.91 | 0.79 | 0.77 | 0.67 | 0.79 | 0.67 |
| $C_i$ | 5 | 4 | 9 | 7 | 7 | 5 | 6 | 9 | 4 | 5 | 6 | 7 | 9 | 8 | 6 |
| $W_i$ | 8 | 9 | 6 | 7 | 8 | 8 | 9 | 6 | 7 | 8 | 9 | 7 | 6 | 5 | 7 |

## VI. PROPOSED METHODOLOGY

We employed IMHS+MDE [47] hybrid for solving the reliability optimization of serial-parallel RRAP presented in the previous section. The hybrid optimization algorithm combines two different techniques: IMHS and MDE. The algorithm works by alternating between running IMHS and MDE for a fixed number of generations. The population from IMHS is passed to MDE and the population from MDE is passed back to IMHS. The process is repeated until reaching the maximum number of iterations or the algorithm converges to an optimal solution. Replacing its selection strategy with that of HS [48], improves DE's exploitation power [49]. With a modified DE selection strategy, the IMHS+MDE hybrid optimization algorithm was used to solve reliability optimization problems in redundancy-reliability allocation when it comes to serial-parallel allocation, and it was effectively efficient.

## VII. RESULTS AND DISCUSSION

Table II presents the effectiveness of the hybrid optimization algorithm. When solving reliability optimization of serial-parallel RRAP, the number of function evaluations (FE) and maximum reliability $R_s$ are taken into consideration. The parameter settings for this hybrid optimization algorithm are CR=0.3, F=1.0, BW=0.5 and PAR=0.2.

Table II compares a variety of different algorithms used for reliability optimization in cloud computing. The score and function evaluations (FE) are what's used to evaluate an algorithm. With these metrics the two that perform the best are IMHS+MDE and IMHS, both with a score of 0.945613 but have 25,890 and 28,377 FE, respectively. Although MGDA and ACO have similar reliability scores their FE is significantly higher. Algorithms INESA, SA, and Luus don't have a failure estimation that was reported and have lower reliability scores. It's important to mention that all of this data depends on the specific parameters and conditions of the optimization problem and may not be useful in other cases. When you're choosing an algorithm, you need to keep in mind things like computational complexity, scalability, and robustness. But one other thing that's also important is the number of redundant components.

TABLE II. COMPARISION

| Algorithm | x1 | x2 | x3 | x4 | x5 | x6 | x7 | x8 | x9 | x10 | x11 | x12 | x13 | x14 | x15 | $R_s$ | FE |
|---|---|---|---|---|---|---|---|---|---|---|---|---|---|---|---|---|---|
| **IMHS+ MDE** [46] | 3 | 4 | 6 | 4 | 3 | 2 | 4 | 5 | 4 | 2 | 3 | 4 | 5 | 4 | 5 | **0.945613** | 25,890[a] |
| **IMHS** [45] | 3 | 4 | 6 | 4 | 3 | 2 | 4 | 5 | 4 | 2 | 3 | 4 | 5 | 4 | 5 | **0.945613** | 28,377[b] |
| **MGDA** [44] | 3 | 4 | 6 | 4 | 3 | 2 | 4 | 5 | 4 | 2 | 3 | 4 | 5 | 4 | 5 | **0.945613** | 217,157 |
| **ACO** [43] | 3 | 4 | 6 | 4 | 3 | 2 | 4 | 5 | 4 | 2 | 3 | 4 | 5 | 4 | 5 | **0.945613** | 244,000 |
| **INESA** [41] | 3 | 4 | 5 | 3 | 3 | 2 | 4 | 5 | 4 | 3 | 3 | 4 | 5 | 5 | 5 | **0.944749** | NA |
| **SA** [41] | 3 | 4 | 5 | 4 | 3 | 2 | 4 | 5 | 4 | 3 | 3 | 4 | 5 | 5 | 4 | **0.943259** | NA |
| **Luus** [41] | 3 | 4 | 5 | 3 | 3 | 2 | 4 | 5 | 4 | 3 | 3 | 4 | 5 | 5 | 5 | **0.944749** | NA |

[a]. The median of 25 runs
[b]. Mean of 30 Runs as given in [45].

## VIII. RESEARCH QUESTIONS OF THE PROPOSED APPROACH

- Simplifying Assumptions: When it comes to reliability optimization in cloud computing, simplifying assumptions are made. The purpose is to make the optimization problem easier. While this makes sense, it's not capturing the full complexity of real-world cloud systems.
- What about the rest?: Sometimes we see research that focuses on something so specific and forgets everything else. Other factors hold just as much importance. This can limit how applicable the results are in real-world cloud systems.
- Real-World Validation: Simulation and theoretical models are the main tools used in reliability optimization studies. And while they do provide insights, they don't reflect all real-world conditions and variability. They just don't hold up.

- Power Hog: Algorithms designed for optimization usually require a lot of power. And when you got many computers running in a system, you best believe it'll heat up fast. The computational complexity puts a cap on how much we can actually optimize.
- Good Here but Not There: Optimizing one cloud system doesn't mean it'll work well with others. This limits how generalizable these results are and hinders our ability to develop widely applicable strategies.

## IX. Future Directions and Conclusions

In conclusion, researchers have found that serial and parallel reliability redundancy allocation is an important field in cloud computing, distributed systems, and critical systems. Allocating redundancy is when extra resources are put into a system to make sure it continues working if something fails. That goal alone holds a lot of weight in terms of improving the availability and fault tolerance of services. This research has led to the development of mathematical models and algorithms that can be used for optimize redundancy allocation. Some factors include the cost of redundancy, how often failures occur, and how much impact failures have on a system or service.

However, there are still many challenges and opportunities for future research in this area. Like adding machine learning techniques to optimize redundancy allocation into dynamic and uncertain environments; creating new techniques for modeling modern systems that have so much complexity; exploring new types of redundancy like software redundancy which can improve how available services are. To better understand the trade-offs between availability, cost, and performance when allocating redundancy in different types of systems and services. Researchers will need to study more in resource-constrained environments such as edge computing and IoT systems. In general, this field has many potential benefits but also drawbacks. It is still very fresh but holds much importance in finding ways to improve the availability and fault tolerance of systems.